\documentclass[12pt]{article}

\usepackage[utf8]{inputenc}
\usepackage{amsmath, amssymb}
\usepackage{graphicx}
\usepackage{hyperref}
\usepackage{booktabs}
\usepackage{cite}
\usepackage{geometry}
\title{Real-Time Band-Grouped Vocal Denoising Using Sigmoid-Driven Ideal Ratio Masking}
\author{Daniel Williams \\ Independent Researcher \\ \texttt{dgwill12@email.asu.edu}}
\date{}

\begin{document}

\maketitle

\begin{abstract}
Real-time, deep learning-based vocal denoising has seen significant progress over the past few years, demonstrating the capability of artificial intelligence in preserving the naturalness of the voice while increasing the signal-to-noise ratio (SNR). However, many deep learning approaches have high amounts of latency and require long frames of context, making them difficult to configure for live applications. To address these challenges, we propose a sigmoid-driven ideal ratio mask trained with a spectral loss to encourage an increased SNR and maximized perceptual quality of the voice. The proposed model uses a band-grouped encoder-decoder architecture with frequency attention and achieves a total latency of less than 10\,ms, with PESQ-WB improvements of 0.21 on stationary noise and 0.12 on nonstationary noise.
\end{abstract}



\section{Introduction}

Live speech enhancement is critical to improving the intelligibility of human voices across many applications. From teleconferencing and phone calls to hearing aids, background noise interferes with a multitude of human-human interactions that occur through a digital medium. Traditional methods such as spectral gating~\cite{boll1979} and noise profiling-based spectral subtraction~\cite{wang2013scaling} have been shown to reduce background noise in real time, but these methods often cause noticeable alterations to the voice and tend to struggle with non-stationary noise.

In recent years, deep learning methods have outperformed traditional denoising methods on both stationary and nonstationary noise. Specifically for removing nonstationary noise, these methods have been able to learn and recognize common patterns found in dynamically noisy environments. However, many state-of-the-art architectures rely on recurrent structures (e.g., GRUs or LSTMs) or large temporal contexts~\cite{tan2018crn, weninger2015lstm, defossez2020}. While effective at removing noise, these designs often introduce excessive latency or require extensive look-ahead frames, violating the 10\,ms threshold generally required for true real-time audio processing~\cite{kates2018osp}.

Apart from latency, one major challenge in audio denoising is increasing SNR while leaving the vocal signal unchanged. Especially in environments with a low SNR and nonstationary noise, models tend to struggle to preserve the quality of the voice while steadily removing the same amount of noise.

To address these issues, we propose a low-parameter causal denoiser capable of handling both stationary and nonstationary noise with high perceptual quality, all while achieving a total latency of less than 10\,ms. The primary contributions of this work are: (1) a sigmoid-driven masking technique that encourages mask values closer to 0 and 1; (2) a spectral loss function tuned for real-time SNR maximization; and (3) a model trained on a vast multilingual corpus covering seven languages and 700+ environments.

\section{Related Work}

Many deep learning approaches employ the U-Net architecture for model efficiency and predictive masking to preserve the naturalness of the voice~\cite{braun2020unet}. These masking models are trained to predict an $M \times N$ matrix corresponding to the $M \times N$ audio segment after the Short-Time Fourier Transform (STFT) is applied. Ideally, regions containing noise in the audio signal would be close to zero in the masking matrix and regions containing the voice would be close to one, so that after applying the Hadamard product the noise is removed while the voice is preserved.

One such mask is the ideal binary mask (IBM)~\cite{cofer2022ibm}, where noisy T-F units are set to 0 and signal-dominant units are set to 1. However, as Yu et al.\ note, this method tends to introduce musical artifacts and residual noise~\cite{yu2020irm}. To address this, the ideal ratio mask (IRM) scales each T-F unit of the noisy signal between 0 and 1 based on the local SNR~\cite{yu2020irm}.

Hasannezhad et al.\ propose a complex ratio mask (CRM), which improves perceptual quality by altering both the phase and magnitude of the noisy audio~\cite{hasannezhad2020crm}. While IBM and IRM operate only on the magnitude, CRM enhances both the magnitude and phase spectra. Williamson et al.\ propose a similar method, emphasizing that enhancing phase is important for improving perceptual quality~\cite{williamson2016crm}. However, despite the perceptual benefits of enhancing both magnitude and phase, such methods significantly increase the number of parameters and computational complexity. Predicting phase is particularly challenging due to its lack of structural regularity. While Hasannezhad et al.\ achieve some reduction in computational time through the use of GRU over LSTM, complex prediction quadruples the computational operations and, at the very least, doubles the model size, making it less suitable for real-time applications~\cite{hasannezhad2020crm}.

\section{Methodology}

\subsection{Signal Representation}

The system processes audio at a sampling rate of 16{,}000\,Hz. To minimize algorithmic latency, the STFT is applied using a window size of 512 samples and a hop length of 64 samples, yielding a frame duration of approximately 4\,ms.

The model runs inference on each individual frame, ensuring that the only latency consists of the 4\,ms wait before the frame can be processed and the computational time of the model itself. To gain temporal context, the model operates on a rolling buffer of 8 past frames, maintaining causality. The input for each frame consists of two channels: (1) 95th percentile-normalized magnitude values clipped between $-10$\,dB and 5\,dB, and (2) phase angles from the noisy STFT.

\subsection{Model Architecture}

We employ an encoder-decoder architecture of approximately 450{,}000 parameters, optimized for low-latency inference. To decrease computational cost, we use band-grouped processing~\cite{luo2020dualpath}, where the 257 total frequency bins are grouped into bands of 8. These bands are processed using Dense layers to capture local frequency relationships within each band, which is more efficient than standard 2D convolutions.

The architecture consists of two encoder stages, a frequency attention bottleneck, and a single decoder, followed by a final Dense layer. Each encoder stage uses residual connections to link local band information to the global signal representation. The frequency attention bottleneck employs a Squeeze-and-Excitation style mechanism: global average pooling is followed by two Dense layers with ReLU and sigmoid activations, directing the model's focus toward informative frequency bands. The decoder employs a U-Net style skip connection to retain fine-grained spectral details that may be lost in the bottleneck~\cite{stoller2018waveunet}. During training, a sigmoid activation on the final Dense layer encourages the model to predict mask values closer to 0 or 1, preserving the voice while suppressing as much noise as possible.

The model is entirely stateless, relying on no internal memory. Temporal information is captured through the 8 past context frames. This architectural choice accelerates inference while still providing sufficient temporal context to handle nonstationary noise.

\subsection{Training Strategy and Loss Functions}

The model is trained to predict an IRM while gaining phase awareness through the 8 frames of context. The IRM defines the relationship between the clean speech magnitude ($|S|$) and the noisy mixture magnitude ($|Y|$) for each T-F bin, clipped between 0 and 1:
\[
M = \text{clip}\left( \frac{|S|}{|Y| + \epsilon},\, 0,\, 1 \right)
\]
where $\epsilon = 10^{-8}$ prevents division by zero.

The total loss $\mathcal{L}_{\text{total}}$ is the weighted sum of three components. First, mask MSE penalizes the direct numerical difference between the predicted mask ($\hat{M}$) and the ground-truth mask ($M$). Second, a log-magnitude loss (weight: 0.3) computes the mean absolute error (L1) between the log-magnitudes of the enhanced speech ($|\hat{S}|$) and clean speech ($|S|$), where $|\hat{S}| = |Y| \cdot \hat{M}$:
\[
\mathcal{L}_{\text{log}} = \bigl|\log(|\hat{S}| + \epsilon) - \log(|S| + \epsilon)\bigr|.
\]
Third, a magnitude L1 loss (weight: 0.2) provides a linear-scale penalty to preserve sharpness of spectral peaks in the vocal signal:
\[
\mathcal{L}_{\text{mag}} = \bigl||\hat{S}| - |S|\bigr|.
\]

To make training more robust, the model was trained on several diverse datasets including the Saraga Carnatic Music Dataset, CommonVoice, Noisy Speech Database, GTSinger, SingingDatabase, VocalSet, and the Acapella Mandarin Singing Dataset. During training, each clean vocal signal was mixed to a random SNR between 5\,dB and 35\,dB with noise audio from the DNS dataset. Additionally, each mixed segment underwent random pitch shifting, random gain shifting, and additive Gaussian noise.

\section{Results and Evaluation}

The proposed real-time denoiser was evaluated using objective speech quality metrics and benchmarked for computational efficiency. Evaluation was performed on both stationary and nonstationary noise environments. Tables~\ref{tab:stationary} and \ref{tab:nonstationary} summarize performance using Perceptual Evaluation of Speech Quality (PESQ) and Short-Time Objective Intelligibility (STOI).

\begin{table}[h!]
\centering
\caption{Objective Metric Evaluation (Stationary Noise)}
\label{tab:stationary}
\begin{tabular}{@{}lccc@{}}
\toprule
\textbf{Metric} & \textbf{Noisy} & \textbf{Enhanced} & \textbf{$\Delta$} \\ \midrule
PESQ-WB         & 1.5522         & 1.7633            & $+$0.2111         \\
PESQ-NB         & 2.5535         & 2.8085            & $+$0.2551         \\
STOI            & 0.8226         & 0.8069            & $-$0.0157         \\
ESTOI           & 0.7139         & 0.7126            & $-$0.0013         \\ \bottomrule
\end{tabular}
\end{table}

\begin{table}[h!]
\centering
\caption{Objective Metric Evaluation (Nonstationary Noise)}
\label{tab:nonstationary}
\begin{tabular}{@{}lccc@{}}
\toprule
\textbf{Metric} & \textbf{Noisy} & \textbf{Enhanced} & \textbf{$\Delta$} \\ \midrule
PESQ-WB         & 1.5760         & 1.6985            & $+$0.1225         \\
PESQ-NB         & 2.1742         & 2.3750            & $+$0.2008         \\
STOI            & 0.7915         & 0.7746            & $-$0.0169         \\
ESTOI           & 0.6695         & 0.6744            & $+$0.0049         \\ \bottomrule
\end{tabular}
\end{table}

The model achieved significant gains in speech quality, with a Wideband PESQ improvement of 0.2111 and a Narrowband PESQ improvement of 0.2551 on stationary noise. While a marginal decrease in STOI was observed ($-$0.0157), the near-zero delta in Extended STOI indicates that the spectral-temporal envelope and intelligibility of the speech were effectively preserved. This trade-off is common in low-latency magnitude-masking models, where noise suppression is prioritized to improve the listening experience. As expected, the model performed better on stationary noise than on nonstationary noise. Nevertheless, the metrics show a substantial increase in quality for nonstationary noise inference as well.

A primary goal of this architecture is real-time inference. Total system latency $L_{\text{total}}$ is defined as the sum of the algorithmic frame delay and the model inference time:
\[
L_{\text{total}} = T_{\text{frame}} + T_{\text{inference}}.
\]
With a hop length of 64 samples at 16{,}000\,Hz, the algorithmic latency is $T_{\text{frame}} = 4.0$\,ms. On a standard CPU inference setup, the model achieved an average $T_{\text{inference}}$ of 2.214\,ms per frame, yielding a total end-to-end latency of \textbf{6.214\,ms}. This is well below the human perceptual threshold for audio-visual synchronization ($\sim$20--40\,ms), making the system suitable for live telecommunications and real-time vocal monitoring.

\section{Conclusion}

The proposed architecture and training strategy produce a fast, low-parameter model capable of suppressing both stationary and nonstationary noise while preserving the vocal signal. The improvements in PESQ demonstrate that replacing recurrent layers with a stateless rolling-buffer context still allows the model to remove a significant amount of nonstationary noise---noise that typically requires longer temporal information to recognize. Additionally, these results demonstrate the viability of magnitude-only enhancement for real-time denoising, showing that the computational overhead of full complex (magnitude + phase) enhancement is not always necessary to achieve meaningful perceptual improvements.


\end{document}